\documentclass[journal=jacsat,manuscript=article]{achemso}

\usepackage[version=3]{mhchem} 
\usepackage{color}
\usepackage{ulem} 
\usepackage{bm} 

\author{Qingchen Yuan}
\affiliation{MOE Key Laboratory of Material Physics and Chemistry under Extraordinary Conditions, and Shaanxi Key Laboratory of Optical Information Technology, School of Science, Northwestern Polytechnical University, Xi'an 710072, China}
\author{Liang Fang}
\affiliation{MOE Key Laboratory of Material Physics and Chemistry under Extraordinary Conditions, and Shaanxi Key Laboratory of Optical Information Technology, School of Science, Northwestern Polytechnical University, Xi'an 710072, China}
\author{Hanlin Fang}
\affiliation{State Key Laboratory of Optoelectronic Materials $\&$ Technologies School of Physics, Sun Yat-sen University, Guangzhou 510275, China}
\author{Juntao Li}
\affiliation{State Key Laboratory of Optoelectronic Materials $\&$ Technologies School of Physics, Sun Yat-sen University, Guangzhou 510275, China}
\author{Tao Wang}
\affiliation{State Key Laboratory of Solidification Processing, Northwestern Polytechnical University, Xi'an 710072, China}
\author{Wanqi Jie}
\affiliation{State Key Laboratory of Solidification Processing, Northwestern Polytechnical University, Xi'an 710072, China}
\author{Jianlin Zhao}
\email{jlzhao@nwpu.edu.cn}
\affiliation{MOE Key Laboratory of Material Physics and Chemistry under Extraordinary Conditions, and Shaanxi Key Laboratory of Optical Information Technology, School of Science, Northwestern Polytechnical University, Xi'an 710072, China}
\author{Xuetao Gan}
\email{xuetaogan@nwpu.edu.cn}
\affiliation{MOE Key Laboratory of Material Physics and Chemistry under Extraordinary Conditions, and Shaanxi Key Laboratory of Optical Information Technology, School of Science, Northwestern Polytechnical University, Xi'an 710072, China}
\title[An \textsf{achemso} demo]
  {Second harmonic and sum-frequency generations from a silicon metasurface integrated with a two-dimensional material}
\abbreviations{IR, NMR, UV}
\begin{document}
\begin{abstract}
Silicon-based nonlinear metasurfaces were implemented only with third-order nonlinearity due to the  crystal centrosymmetry and the efficiencies are considerably low, which hinders their practical applications with low-power lasers. Here, we propose to integrate a two-dimensional GaSe flake onto a silicon metasurface to assist high-efficiency second-order nonlinear processes, including second-harmonic generation (SHG) and sum-frequency generation (SFG). By resonantly pumping the integrated GaSe-metasurface, which supports a Fano resonance, the obtained SHG is about two-orders of magnitude stronger than the third-harmonic generation from the bare silicon metasurface. In addition, thanks to the resonant field enhancement and GaSe's strong second-order nonlinearity, SHG of the integrated structure could be excited successfully with a low-power continuous-wave laser, which makes it possible to further implement SFG. The high-efficiency second-order nonlinear processes assisted by  two-dimensional  materials present potentials to expand silicon metasurface's functionalities in nonlinear regime. 
\end{abstract}
\section{Introduction}
The recently developed all-dielectric metasurfaces provide a sophisticated platform for controlling light-matter interactions, which reached remarkable efficiencies matching or out-performing conventional optical elements~\cite{DOI: 10.1021/acsphotonics.7b01217,Optica.4.814.2017,DOI: 10.1021/acsphotonics.7b01608,DOI: 10.1021/acsphotonics.8b00098,Nano Lett. 2016.16.4396-4403,Sci. Rep. 2.492.2012,Science.352.1190.2016,Kivshar2018,Kivshar2018-2}. Comparing to metal metasurfaces, the all-dielectric ones have no Ohmic losses, promising a high laser damage threshold. Also, optical fields are localized in the dielectrics for more effective light-matter interactions. Therein, silicon has emerged as one of the most promising materials for dielectric metasurfaces. It has large linear and nonlinear optical susceptibilities to guarantee well-confined electric and magnetic resonances and distinct nonlinear processes~\cite{DOI: 10.1038/NPHOTON.2017.39,Nature Nano.10.937.2015,Science.345.298.2014}. On the other hand, its wafer-growth is mature and fabrication is compatible with the low-cost complementary-metal-oxide-semiconductor (CMOS) technology. In linear optical regime, comprehensive manipulations of amplitude, phase, and polarization of light beams have been successfully carried out in silicon metasurfaces~\cite{Nature Nano.10.937.2015,Science.345.298.2014}, which give rise to ultrathin gratings and axicons~\cite{Science.345.298.2014}, lenses with near-unity numerical aperture~\cite{DOI: 10.1021/acs.nanolett.8b00368}, and compact holograms with diffraction efficiencies exceeding 90\%~\cite{Nano Lett. 2015.15.6261,ACS Photonics 2016.3. 514,ACS Photonics 2017.4.544}. 

Moreover, silicon's nonlinear optical responses have been widely investigated in third-harmonic generations (THGs), four-wave mixings (FWMs), and all-optical modulations~\cite{Shen Y. R. The Principles of Nonlinear Optics (Wiley 1984), Boyd R. W. Nonlinear Optics 3rd edn (Academic 2008)}. If these optical frequency generations or intensity-dependent refractive index were involved into silicon metasurfaces, their above linear optical functionalities could be further greatly reinforced~\cite{oi:10.1038/natrevmats.2017.10}. For instance, by simultaneously exciting THGs of electric and magnetic Mie multipoles in silicon metasurfaces, directional deflections and optical vortex beams with orbital angular momentum could be realized via the smooth phase gradients of the generated third-harmonic field~\cite{DOI: 10.1021/acs.nanolett.8b01460,DOI: 10.1021/acsphotonics.7b01277}. THGs, FWMs and all-optical Kerr effects in silicon metasurfaces have been reported to have efficiencies several orders of magnitude higher than those of the non-structured silicon slab~\cite{DOI: 10.1021/acsphotonics.7b01423,Phil. Trans. R. Soc. A 375: 20160281,Nano Lett. 2015 15 6985,Nano Lett. 2016 16 4857,Nano Lett. 2014 14 6488,DOI: 10.1021/acs.nanolett.5b02802}. However, the absolute values of these efficiencies are still very low (in the orders of 10$^{-6}$ with a power density of GW/cm$^2$), challenging the realistic applications of nonlinear silicon metasurfaces. In nonlinear optics, to realize high-efficiency harmonic generations or electro-optical modulations, second-order nonlinear responses are good alternatives due to the much higher coefficients ($\chi^{(2)}$) than the third-order nonlinear coefficients ($\chi^{(3)}$). Unfortunately, limited by silicon's crystal centrosymmetry, it has no second-order nonlinearity to realize high-efficiency nonlinear processes in its metasurface. 

In this paper, we propose and demonstrate, by integrating a two-dimensional gallium selenide (GaSe) flake onto a silicon metasurface, it is possible to carry out high-efficiency second-order nonlinear processes from the silicon metasurface. Pumping the integrated GaSe-metasurface with an on-resonance laser, the frequency up-conversion spectrum presents a strong second-harmonic generation (SHG) signal, which could be two orders of magnitude stronger than the THG signal of the bare silicon metasurface. We ascribe this high SHG efficiency to the large $\chi^{(2)}$ of GaSe as well as the near-field enhancement of the metasurface. Hence this field-enhancement of the integrated GaSe-metasurface even supports the successful excitation of SHG using a continuous-wave (CW) laser, which makes it possible to further implement sum-frequency generation (SFG) without synchronized multiple pulsed lasers.  

Two-dimensional GaSe is chosen because it has high $\chi^{(2)}$. Its monolayer primary was reported to have a $\chi^{(2)}$ as $\sim$1,000 pm/V with the pump laser at the telecom-band~\cite{J. Am. Chem. Soc. 2015 137 7994}, which is about two orders of magnitude higher than those in conventional bulk materials. In addition, with the stacking order of $\varepsilon$-GaSe, its few-layers are always absent of centrosymmetry. So that the second-order nonlinear processes would be additive with the layer numbers. Meanwhile, as one type of two-dimensional materials, GaSe has advantages for constructing integrated optoelectronic devices with nanophotonic structures~\cite{J. Am. Chem. Soc. 2015 137 7994}. For instance, its flexibility allows the easy integration with nanophotonic structures in micrometer scale~\cite{Gan_LightSA}. Furthermore, as the development of chemical vapor deposition growth technique, it is predictable to grow two-dimensional materials onto silicon wafers directly for large-scale fabrication of optoelectronic devices.

\section{Device fabrication}
 \begin{figure}[th!]\centering
	\includegraphics[width=5.0in]{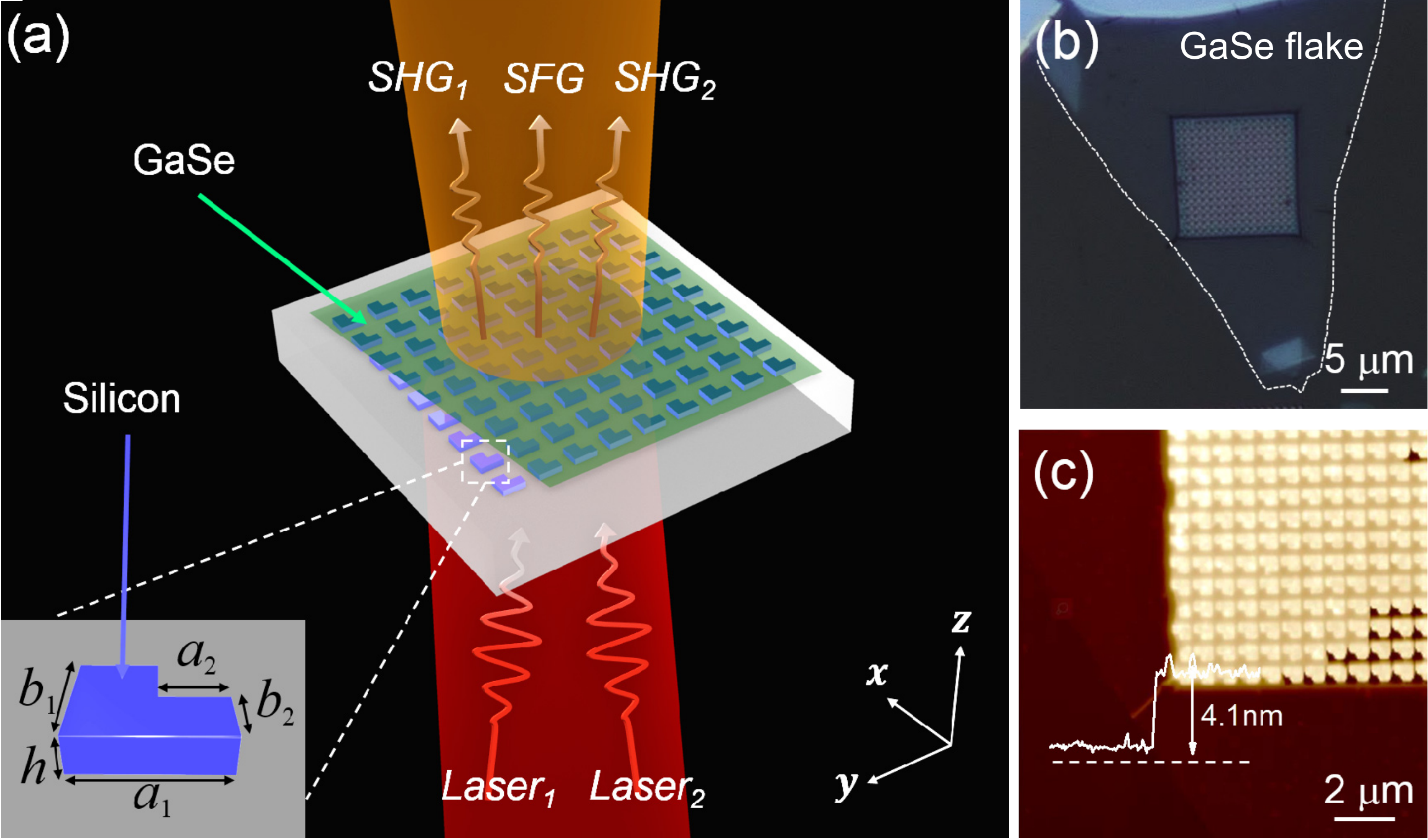}
	\caption{{\small (a) Schematic of operations of SHG and SFG from the GaSe-integrated silicon metasurface. Inset: Structure of the $L$-shaped asymmetric meta-atom. (b) Optical microscope image of the fabricated device, where the white dashed line shows the profile of integrated GaSe flake. (c) AFM image of the fabricated device, showing the GaSe thickness of 4.1 nm.}}
	\label{struc} 
\end{figure}

Figure~\ref{struc}(a) schematically displays the operations of nonlinear processes in the GaSe-integrated silicon metasurface. Here, the silicon metasurface is designed with an array of $L$-shaped asymmetric meta-atoms~\cite{ACS Photonics 2016 3 2362}. Because of the coupling between the bright in-plane electric dipole mode and the dark longitudinal magnetic dipole mode, this metasurface supports Fano resonances with high quality ($Q$) factors. With the pumps of two on-resonance lasers, besides the intrinsic THGs originated from the $\chi^{(3)}$ of the silicon metasurface, there are expected SHG and SFG assisted by the two-dimensional GaSe flake. The linear and nonlinear optical characterizations of the integrated device are implemented using an optical transmission system. The excitation laser is focused onto the metasurface pattern using an objective lens with a numerical aperture of 0.42. After the sample, there is another objective lens with a numerical aperture of 0.42 to collect the linear transmissions and the generated frequency up-conversion signals, which are then divided by a dichroic mirror (cutoff wavelength of 1,000 nm). Only for the SFG process, two pump lasers are employed simultaneously and combined by a beam splitter in the input optical path. Otherwise, only one pump laser is required.

A 230 nm thick silicon membrane grown on a sapphire substrate is used to fabricate the metasurface, with the processes of electron beam lithography and plasma dry etching. The specific parameters of a single meta-atom are labelled in the inset of Fig.~\ref{struc}(a), which are $a_1$=550 nm, $a_2$=306 nm, $b_1$=428 nm, $b_2$=$b_1$/2, $h$=230 nm. The lattice period of the metasurface is  $p_x$=$p_y$= 830 nm, which supports Fano resonance mode in the telecom-band.  GaSe flake is prepared on polydimethylsiloxane by mechanical exfoliation from the bulk material, which is then dry transferred onto the silicon metasurface~\cite{transfer}. Figure \ref{struc}(b) shows an optical microscope image of the fabricated device. Because GaSe has a larger refractive index than that of sapphire, its flake shows different color from the bare substrate. The thickness of the GaSe layer is examined using an atomic force microscope (AFM), as shown in Fig.~\ref{struc}(c). The boundary of the flake shows a thickness of 4.1 nm, corresponding to a five-layer considering a monolayer thickness of 0.8 nm. Moreover, the morphology over the silicon metasurface indicates the tight contact of the GaSe flake. 

\section{Results and discussions}
\begin{figure}[th!]\centering
	\includegraphics[width=4.5in]{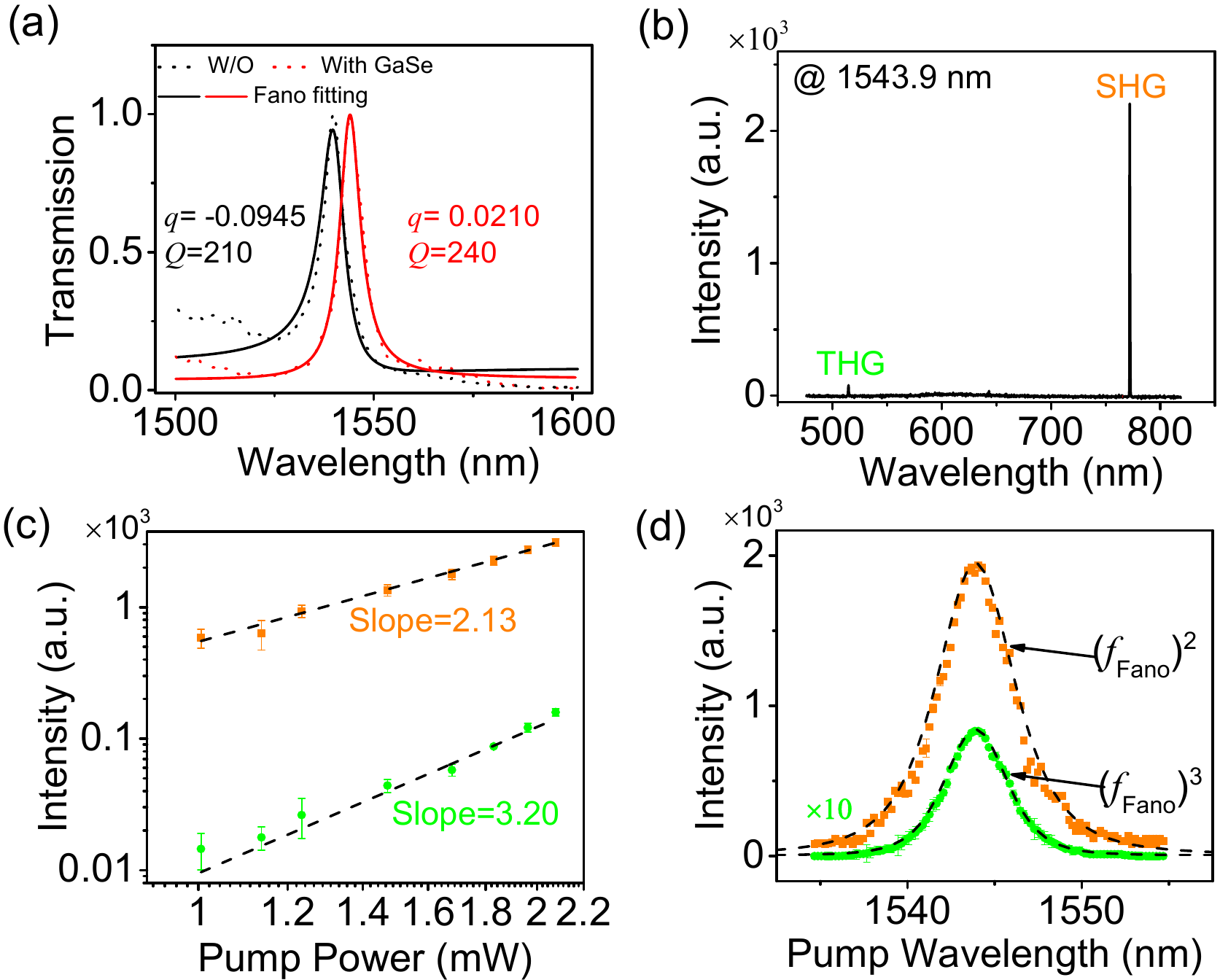}
	\caption{{\small (a) Measured transmission spectra of the silicon metasurface without and with GaSe nanoflake as well as their fittings of Fano functions. (b) Frequency up-conversion spectrum of the integrated GaSe-metasurface pumped at 1543.9 nm, showing SHG and THG peaks at 771.9 nm and 514.6 nm, respectively. (c) Log-log plots of SHG and THG's power-dependences on the pump powers. (d) Pump wavelength-dependences of SHG and THG when the pump wavelength is turned across the resonant wavelength, where \textit{f}$_{Fano}$ is the Fano function used to fit the resonant mode in (a).}}
	\label{shg} 
\end{figure}

To characterize the resonance mode of the silicon metasurface, a narrowband tunable continuous-wave (CW) laser is focused onto the sample, and the transmitted optical powers are monitored using a telecom-band photodiode. By tuning the wavelength of the incident laser, the transmission spectrum could be obtained. The resonant mode of the broken-symmetry metasurface has a linearly polarized far-field radiation following the radiation formulation~\cite{ACS Photonics 2016 3 2362}. To reach a high mode-coupling efficiency, the laser polarization is optimized to match the mode polarization, which is along the $x$-axis labelled in Fig.~\ref{struc}(a). Figure \ref{shg}(a) displays the measured transmission spectra from the silicon metasurface before and after integrating five-layer GaSe flake. Asymmetric Fano peak and dip are obtained due to the constructive and destructive interferences between the electric and magnetic resonant dipoles. They could be fitted by the Fano function $f_{Fano}$=A$_0$+F$_0$($q$+2($\omega$-$\omega$$_0$)/$\varGamma$)$^2$/(1+(2($\omega$-$\omega$$_0$)/$\varGamma$)$^2$), where $q$ is the Fano factor, meaning the energy ratio of  discrete state and continuous state, $\omega_0$ is  the resonant frequency, $\varGamma$ is the width of resonant mode, and $A_0$ and $F_0$ are constants. The peak wavelength undergoes a  red-shift from 1540.9 nm to 1543.9 nm, resulting from the increased dielectric functions around the resonant mode with the integration of GaSe. Meantime, the $Q$ factor has a slightly increased value from 210 to 240. It could be attributed to the improved symmetry of the refractive index below and above the metasurface, which ensures better confinement of the vertical total internal reflection. To confirm the experimental results, mode simulations are carried out using a finite element technique (COMSOL Multiphysics). The GaSe-induced variations of the peak wavelength and $Q$ factor are similar as those obtained in the experiments. Moreover, the optical power confined in the GaSe flake is estimated as 3.06\% of that in the silicon metasurface. The mode distributions of the GaSe-integrated and bare metasurfaces indicate, while there are optical fields in the GaSe flake, the optical field distributes much better inside the metasurface after the integration of GaSe, which responds to the increased $Q$ factor.

Fano resonance modes in metasurfaces could notably provide strong localized optical fields to enhance light-matter interaction. To excite the enhanced harmonic generations from the Fano resonance of the integrated GaSe-metasurface, a tunable pulsed laser is employed with a pulse width of 8.8 ps and a repetition rate of 18.5 MHz. By tuning the laser wavelength to match the peak wavelength of the Fano resonance ($\sim$1543.9 nm ), the frequency up-conversion signals are acquired using a spectrometer mounted with a cooled silicon camera. Figure~\ref{shg}(b) shows a measured spectrum with a pump power of 1.65 mW, which is measured before the GaSe-metasurface. At the wavelengths of 514.6 nm and 771.9 nm, corresponding to the wavelengths of THG and SHG of the pump laser, there are two peaks. To verify these harmonic processes, the pump power is varied gradually, and powers of the SHG and THG are recorded, as shown in Fig.~\ref{shg}(c). These pump power dependences are linearly fitted with slopes of 2.13 and 3.20 in log-log coordinate system, respectively, which are typical characteristics of SHG and THG. To study origins of the SHG and THG signals, we implement the same measurements of harmonic generations from the bare silicon metasurface before the integration of GaSe flake. By resonantly pumping the metasurface with a pulsed laser at 1540.9 nm, only a weak THG peak is obtained, arising from silicon's intrinsic third-order nonlinearity. There is no detectable SHG from the bare silicon metasurface due to silicon's inversion centrosymmetry. Even if silicon surface has the broken centrosymmetry, the supported SHG should be two orders of magnitude weaker than the bulk THG~\cite{Opt. Express 2010 18 26613}. Hence, the obtained strong SHG of the GaSe-metasurface arises from the strong second-order nonlinearity of the integrated GaSe. At the pump power of 1 mW, the ratio of the SHG and THG powers is about 101. While this ratio would be reduced for even higher pump powers, the SHG realized by the GaSe nanoflake is much stronger than the THG obtained from the 230 nm silicon metasurface. It illustrates GaSe's assistance on the second-order nonlinear processes of silicon metasurfaces. 

SHG and THG signals from the integrated GaSe-metasurface are further studied by tuning the wavelength of the pulsed pump laser, as plotted in Fig.~\ref{shg}(d). As the pump wavelength is tuned away from the resonant wavelength (1543.9 nm), both of the SHG and THG intensities decrease gradually and eventually reduce to undetectable values. It implies generations of harmonic signals rely on the enhancement effect of the metasurface's resonant mode. For the incident laser with varied wavelengths coupled with the integrated GaSe-metasurface, the optical power confined in the resonant mode is governed by the Fano function $f_{Fano}$, which is extracted from the fitting curve in Fig.~\ref{shg}(a) (the solid red line). In the processes of SHG and THG, the generated optical powers are quadratic and cubic functions of the pump power. Therefore, as shown in Fig.~\ref{shg}(d), the experimental results of SHG and THG are fitted well by $(f_{Fano})^2$ and $(f_{Fano})^3$, respectively.

\begin{figure}[th!]\centering
	\includegraphics[width=4.5in]{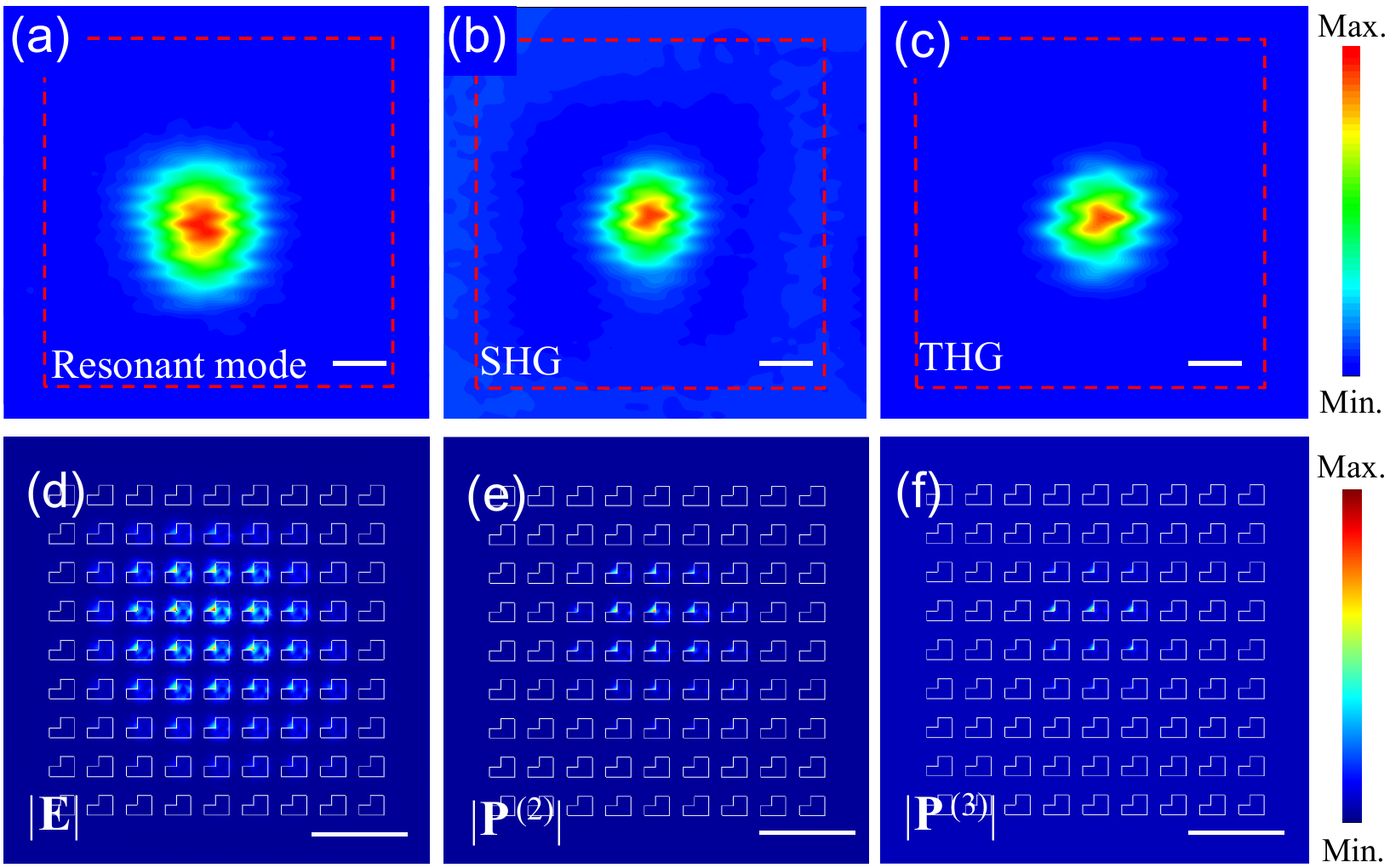}
	\caption{{\small (a)-(c) Spatial mappings of (a) the resonant mode and the corresponding (b) SHG and (c) THG. The red dashed box implies the boundary of the metasurface. (d)-(f) Simulated near-field distributions of (d) $\left|\textbf{E}\right|$, (e)$\left|\textbf{P}^{(2)}\right|$ and (f) $\left|\textbf{P}^{(3)}\right|$. Scale bar is 2 $\mu$m.}}
	\label{map} 
\end{figure}

The afore-obtained resonantly enhanced SHG and THG are further illustrated by implementing their spatial mappings. Pumped with the on-resonance pulsed laser, the device is mounted on a two-dimensional piezo-actuated stage with moving steps of 300 nm. Transmission of the resonant mode and far-field radiations of the SHG and THG signals are monitored using near-infrared and visible grating spectrometers. Figures \ref{map}(a)-(c) display the corresponding measurement results. To elucidate the experiment results, the mode distribution ($\left|\textbf{E}\right|$) of the integrated GaSe-metasurface is simulated, as shown in Fig.~\ref{map}(d). As mentioned above, the employed Fano resonance mode with a high $Q$ factor arises from couplings of electric and magnetic dipoles between the symmetry-broken meta-atoms. For the meta-atoms around the boundary, there is no mode confinement due to the radiative dipole along the $z$-direction. Hence, the optical field only localizes around the metasurface center. Also, because of the symmetry-breaking of the meta-atoms, the mode distribution is not circularly symmetric. Only when the focused incident laser overlaps with the mode location of the metasurface, the resonant mode could be excited successfully. Hence, in the spatial mapping of the mode distribution, the effective transmissions are only observed at the center, which are in well agreement with the experimental results. The measured SHG and THG are excited by the near-field of the Fano resonance, and their radiations are determined by the nonlinear polarizations induced in the GaSe flake or silicon metasurface. With the simulated electrical field distributions (\textbf{E}) of the resonant mode, the corresponding nonlinear polarizations could be calculated from
\begin{equation}
\textbf{P}^{NL}=\textbf{P}^{(2)}+\textbf{P}^{(3)}+\cdots=\varepsilon_0\bm{\chi}^{(2)}:\textbf{E}\textbf{E}+\varepsilon_0\bm{\chi}^{(3)} \vdots \textbf{E}\textbf{E}\textbf{E}+\cdots
\end{equation}
Figures \ref{map}(e) and (f) display the near-field distributions of $ \left|\textbf{P}^{(2)}\right|$ and $ \left|\textbf{P}^{(3)}\right|$ for SHG and THG, which have smaller areas than that of the resonant mode. When the objective lens collects the SHG and THG signals, their spatial mappings of far-field radiations are determined by the near-field distributions of $\left|\textbf{P}^{(2)}\right|$ and $\left|\textbf{P}^{(3)}\right|$, which agree well with the experimental results shown in Figs.~\ref{map}(b) and (c).

In the reported nonlinear processes in silicon metasurfaces, caused by silicon's weak nonlinearity,  it is essential to exploit a pulsed laser as the excitation to provide an ultrahigh peak intensity. On the other hand, most of SHGs in two-dimensional materials were implemented using a pulsed laser as well, because of the weak light-matter coupling in the atomically thin layer. Here, in the integrated GaSe-metasurface, we further demonstrate it is possible to realize the CW operations of second-order nonlinear processes. It benefits from the near-field enhancement of metasurface's resonance mode as well as the ultrastrong $\chi^{(2)}$ in two-dimensional GaSe. By changing the excitation from an on-resonance pulsed laser to an on-resonance CW laser, SHG is obtained from the integrated structure. With the pump power of 14.61 mW focused on the device, the SHG signal with a power of 5.01 nW is measured using a visible photomultiplier tube, corresponding to a conversion efficiency of $\eta_{SHG}$= $\frac{P_{SHG}}{P_{Pump}^2}$=2.477$\times$10$^{-5}$ W$^{-1}$.
\begin{figure}[th!]\centering
	\includegraphics[width=6.2in]{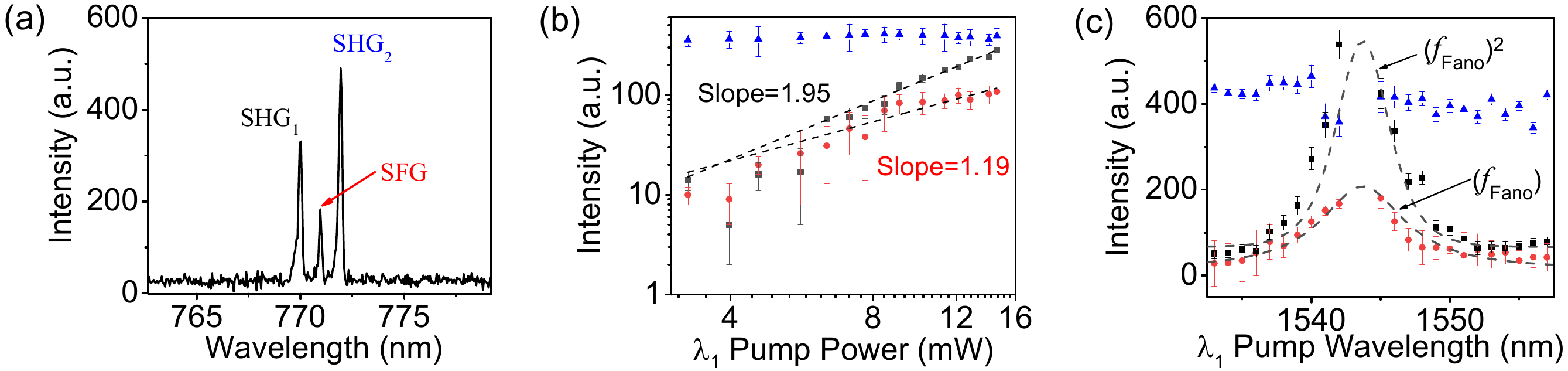}
	\caption{{\small (a) Frequency up-conversion spectrum obtained from the GaSe-metasurface when pumped by two CW lasers at the wavelengths of 1540.0 nm ($Laser_1$) and 1543.9 nm ($Laser_2$), showing SHG and SFG peaks. (b) Log-log plots of power-dependences of SHG and SFG on the pump power of $Laser_1$. (c) Pump wavelength-dependences of SHG and SFG when the wavelength of $Laser_1$ is turned across the resonant wavelength and the wavelength of $Laser_2$ is maintained, where the red (black) line is a \textit{f}$_{Fano}$ ((\textit{f}$_{Fano}$$)^2$) fitting curve.}}
	\label{sfg} 
\end{figure}

The CW-pumped SHG in the integrated device indicates it is commodious to implement other second-order nonlinear processes thanks to the exemption of synchronization of multiple pulsed lasers~\cite{AOM_multipleOFC}. To carry out that, we input two CW lasers ($Laser_1$ and $Laser_2$) simultaneously as the excitations. Their wavelengths are tuned as 1540.0 nm ($Laser_1$) and 1543.9 nm ($Laser_2$), which locate within the Fano resonance peak to on-resonantly excite the resonant modes. Figure~\ref{sfg}(a) displays the measured frequency up-conversion spectrum with the two CW on-resonance pumps. There are three peaks locating at the wavelengths of 770.0 nm, 771.0 nm, and 771.9 nm. According to the wavelength conversion, they are recognized as the SHGs and SFG of the two pump lasers. To facilitate the discussion, the abovementioned three peaks are labelled  as $SHG_1$, $SFG$, and $SHG_2$ successively. By fixing the pump power of $Laser_2$ and varying the pump power of $Laser_1$ gradually, the powers of the three frequency up-conversion peaks are acquired, as shown in Fig.~\ref{sfg}(b). Similar as that shown in Fig.~\ref{shg}, in the SHG of $Laser_1$, two photons of this fundamental wave convert into one photon of the frequency up-conversion wave. The quadratic function (slope=1.95) of the pump power is obtained for this process. In the process of SFG, where both of the two pump lasers involve, to generate one frequency up-converted photon, the $Laser_1$ contributes one photon, resulting in a linear power-dependence (slope=1.19). 

The possibility of SFG with two CW pumps arises from the coupling between GaSe and the high density of electrical field in the Fano resonance mode. To prove that, we examine SFG's dependence on the pump wavelength. The wavelength of $Laser_2$ is fixed (1543.9 nm), and the wavelength of $Laser_1$ is scanned across the whole resonant peak range. The measured SFG powers are plotted in Fig.~\ref{sfg}(c). For the laser wavelengths detuned away from the Fano resonant peak, SFG signals decrease to undetectable levels due to the weak light-GaSe coupling in the single-passed normal radiation. When the laser wavelength is tuned into the resonant peak, the resonant mode is excited and the densities of photon states at different wavelengths would be described by the Fano function $f_{Fano}$ obtained from the fitting of Fig. \ref{shg}(a). Therefore, when the pump laser is scanned across the resonant mode, the SFG power would vary in the form of $f_{Fano}$, as indicated by the fitting curve in Fig.~\ref{sfg}(c). The wavelength dependence of  $SHG_1$ is plotted in Fig.~\ref{sfg}(c) as well, showing a fitting function of $(f_{Fano})^2$ considering its quadratic function of the pump power of $Laser_1$. 

\section{Conclusion}
In conclusion, we have demonstrated the realizations of second-order nonlinear processes, including SHG and SFG, from a silicon metasurface with the assistance of a two-dimensional GaSe flake. Benefitting from GaSe's strong second-order nonlinearity, the resonantly pumped GaSe-metasurface presents strong SHG that is two orders of magnitude higher than THG of the bare silicon metasurface. In addition, the high densities of electrical field in the Fano resonance mode enable the successful excitation of CW-pumped SHG from the GaSe-metasurface, and the conversion efficiency is estimated as 2.477$\times$10$^{-5}$ W$^{-1}$. The CW operations of second-order nonlinear processes implies the exemption of synchronization of multiple pulsed lasers, which allows an easy implementation of SFG as well. The demonstrated  high-efficiency second-order nonlinear processes assisted by two-dimensional materials could provide a routing to developping various nonlinear functionalities on the basis of silicon metasurfaces~\cite{oi:10.1038/natrevmats.2017.10,DOI: 10.1021/acs.nanolett.8b01460,DOI: 10.1021/acsphotonics.7b01277}.

\begin{acknowledgement}
	Financial support was provided by National Natural Science Foundations of China (61775183, 11634010); the Key Research and Development Program (2017YFA0303800); the Key Research and Development Program in Shaanxi Province of China (2017KJXX-12, 2018JM1058); the Fundamental Research Funds for the Central Universities (3102017jc01001, 3102018jcc034, 3102019JC008), and the Innovation Foundation for Doctor Dissertation of Northwestern Polytechnical University (CX201924). We would like to thank the Analytical $\&$ Testing Center of NPU for the assistances of device fabrication.
\end{acknowledgement}

\end{document}